\documentclass[pdflatex,iicol,sn-mathphys-num]{sn-jnl}

\usepackage{amsmath,amssymb}
\usepackage{graphicx}
\usepackage{booktabs}
\usepackage{subcaption}
\usepackage{placeins}
\usepackage{tabularx}
\providecommand{\burl}[1]{\url{#1}}

\begin{document}

\title{Separating charm and bottom decay electrons with point-cloud deep learning on the charged-hadron environment}

\author[1]{\fnm{J. Y.} \sur{Zhang}}
\author*[1]{\fnm{W. B.} \sur{He}}
\email{hewanbing@fudan.edu.cn}
\author*[1]{\fnm{L.} \sur{Ma}}
\email{malong@fudan.edu.cn}

\affil*[1]{Key Laboratory of Nuclear Physics and Ion-beam Application (MOE), Institute of Modern Physics, Fudan University, Shanghai 200433, China}

\abstract{
Electrons from semi-leptonic decays of charm ($D$) and bottom ($B$) hadrons are important probes in high-energy collisions, while their separation remains challenging due to the similarity of the underlying decay topologies. In this work, the charged-hadron environment associated with each electron is represented as an unordered point cloud and combined with the electron kinematics and charge. Several set-based machine learning architectures, including DeepSets, Transformer, and graph neural networks, are used to distinguish electrons from prompt $D$ and direct $B$ decays without explicit vertex observables. The three architectures achieve comparable performance. Moreover, the Transformer is found to perform better than a boosted decision tree trained on physics-motivated high-level observables. At an efficiency of about 40\%, the classifier reaches a purity level close to 80\% on the test sample. The classifier response is strongly related to the geometric and topological properties of the charged-hadron environment. The information learned by the classifier is further studied using non-prompt $D$ electrons and charged hadrons from different angular regions. Our results indicate that the learned separation is driven mainly by the later stages of the decay chain rather than by initial production effects. The angular-region study further shows that useful discriminating information extends beyond the near- and away-side cones. This method is also evaluated for mid-rapidity proton--proton collisions at both RHIC and LHC energies.
}

\maketitle

\section{Introduction}

Heavy-flavor observables have been widely used as important probes in high-energy collisions~\cite{Rapp:2018qla, Dong:2019unq, He:2022ywp, Zhang:2025tgt, Zhao:2020jqu}. Electrons from semi-leptonic decays of heavy-flavor hadrons provide important probes for studying heavy-flavor production and medium properties in high-energy collisions~\cite{ALICE:2018gyx, Masciocchi:2011fu, ALICE:2014ivb, ALICE:2012acz, Averbeck:2004dv, PHENIX:2010xji}. In particular, separating electrons originating from charm ($D$) and bottom ($B$) hadrons is essential for many heavy-flavor measurements. However, this separation remains challenging because charm- and bottom-decay electrons exhibit similar kinematic characteristics.

To achieve such a separation, a variety of experimental techniques have been developed. Existing experimental approaches mainly rely on displaced-vertex information~\cite{PHENIX:2015ynp, ALICE:2019nuy, ALICE:2021mgk, STAR:2021uzu, ALICE:2019bfx}, such as the distance of closest approach (DCA), to distinguish charm and bottom contributions. These methods primarily exploit the lifetime difference between charm and bottom hadrons and have achieved considerable success experimentally. Nevertheless, they do not fully exploit correlations encoded in the hadronic environment. Differences in heavy-flavor production, fragmentation, and decay processes can lead to different patterns of associated hadronic activity around charm- and bottom-origin electrons~\cite{ALICE:2023kjg, Thomas:2024cso, Andronic:2015wma, PHENIX:2009dpd, ALICE:BeautyElectronHadron2014}.

This raises the question of how much information about the heavy-flavor origin is encoded in the hadronic environment itself, independent of explicit vertex observables. Traditional approaches typically characterize the hadronic activity using a limited number of physics-motivated high-level observables describing the local hadronic structure around the electron. While such observables are physically motivated and provide effective low-dimensional descriptions of the hadronic environment, they may not fully exploit the complete information encoded in the multi-particle event structure~\cite{Komiske:2018cqr}. In addition, combining many correlated observables in a systematic way becomes increasingly difficult as the complexity of the event structure grows~\cite{Albertsson:2018maf}.

Recent advances in machine learning have made it possible to analyze high-dimensional particle-level data directly, enabling more complete exploitation of the information encoded in complex event structures~\cite{Karagiorgi:2021ngt, Mondal:2024nsa, Zaheer:2017wmg, Wang:2018nkf, lee2019set, Carleo:2019ptp, CMS:HeavyFlavorJetTagging2018, Guest:JetFlavorDNN2016, Bols:DeepJet2020, gao2021machine}. In high-energy physics, collider events can be naturally represented as unordered particle sets, or point clouds, making set-based architectures particularly suitable for particle-level event modeling~\cite{Albertsson:2018maf, Kasieczka:2021xcg, Guest:2018yhq, Radovic:2018dip, BelleII:GFlaT2024, LHCb:InclusiveFlavorTagging2025}. Architectures such as DeepSets, Transformer-based models, and graph neural networks have demonstrated strong performance in jet tagging, event classification, and jet substructure studies~\cite{Louppe:2017ipp, Komiske:2018cqr, Qu:2019gqs, Qu:2022mxj, Zhu:2023xpk, Shlomi:GNNParticlePhysics2021, ATLAS:GN2FlavorTagging2026, he2023high}.

In this work, the charged-hadron environment associated with each electron is represented as an unordered point cloud and combined with the electron $p_T^e$, $\eta^e$, and charge. We study how effectively electrons from prompt $D$ and direct $B$ decays can be separated without explicit vertex observables. Three set-based models, DeepSets, Transformer, and graph neural networks(GNN), are systematically compared. The relationship between the classifier response and physics-motivated observables are studied, together with feature perturbations, to identify the properties of the charged-hadron environment that are most relevant for the separation. The learned information is further examined using non-prompt D electrons and charged hadrons selected from different angular regions.

\section{Methodology}

\subsection{Dataset and Representation}
\label{sec:dataset}

The datasets used in this study are generated with \textsc{Pythia}~8.312 using the Monash 2013 tune~\cite{PythiaReference,Skands:Monash2013}. The main classification and interpretability studies are performed at $\sqrt{s}=200$ GeV. Heavy-flavor electron candidates of both charge signs are selected in the transverse-momentum range $3<p_T^e<10~\mathrm{GeV}/c$. The 200-GeV samples use $|\eta^e|\leq 5$ for electrons and select charged hadrons with $p_T^h\geq 0.2~\mathrm{GeV}/c$, $|\eta^h|\leq 5$, and full azimuth. Electrons from prompt $D$-hadron decays and direct $B$-hadron decays form the two classification classes. Non-prompt $D$ electrons from $H_b\rightarrow D\rightarrow e$, where $H_b$ denotes a bottom hadron, are kept separate from the binary training sample.

\begin{figure}[htb]
    \centering
    \includegraphics[width=\linewidth]{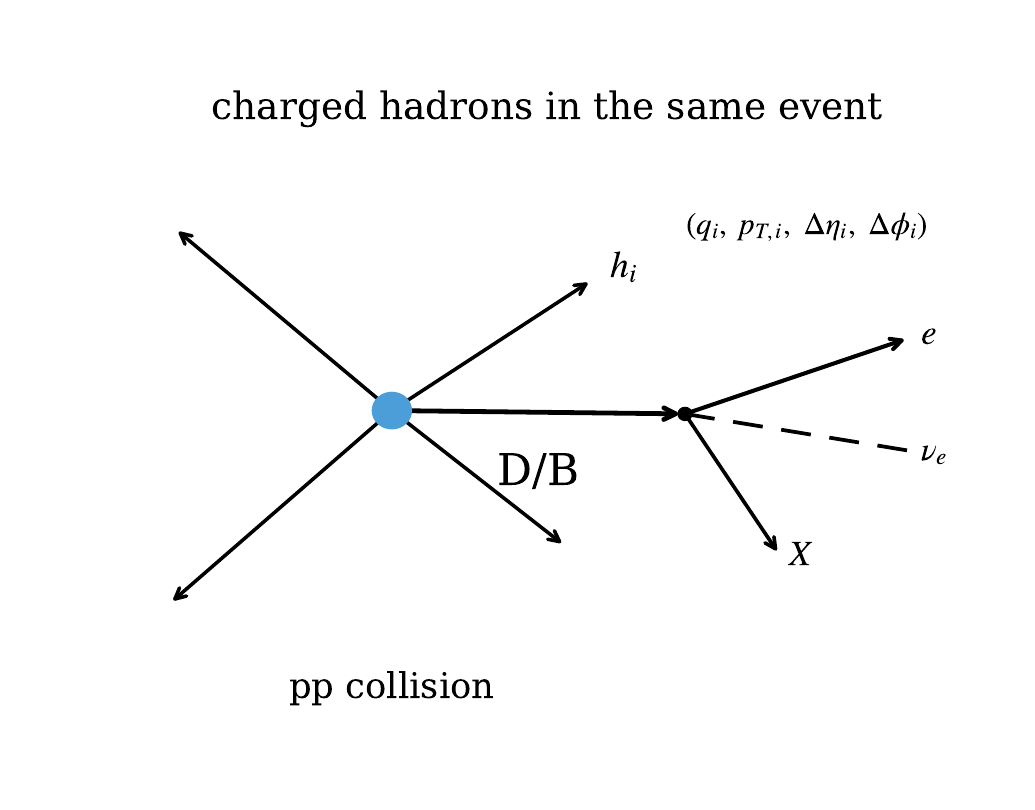}
    \caption{Schematic of a heavy-flavor decay electron and charged hadrons in the same event.}
    \label{fig:dataset_generator_schematic}
\end{figure}

For each selected electron, an event-level sample is constructed by associating the electron with final-state charged hadrons in the same event, as illustrated in Fig.~\ref{fig:dataset_generator_schematic}. The hadrons are selected within the stated acceptance without an angular-cone restriction. Each hadron is characterized by its transverse momentum $p_{T,i}$, angular displacement relative to the electron $(\Delta\eta_i,\Delta\phi_i)$, and electric charge $q_i$.

For the model input, the azimuthal displacement is represented by $\sin\Delta\phi_i$ and $\cos\Delta\phi_i$ to avoid the discontinuity at the azimuthal boundary. The hadronic environment is represented as an unordered set of hadron feature vectors and combined with the electron transverse momentum, pseudorapidity, and charge. The particle selections and input representation for the 200 GeV proton collision studies are summarized in Table~\ref{tab:dataset_representation}.

\begin{table}[htb]
    \caption{Dataset construction and model inputs.}
    \label{tab:dataset_representation}
    \centering
\small
\setlength{\tabcolsep}{4pt}
\renewcommand{\arraystretch}{1.20}

\begin{tabularx}{\columnwidth}{@{}
    >{\raggedright\arraybackslash}p{0.20\columnwidth}
    >{\raggedright\arraybackslash}X@{}}
    \toprule
    \multicolumn{2}{@{}l}{\textbf{Objects and selections}} \\[2pt]

    HF $e^{\pm}$
      & Electron from a semileptonic HF decay:\newline
        $3\leq p_T^e<10~\mathrm{GeV}/c$, $|\eta_e|\leq5$. \\[3pt]

    Hadron select
      & Charged hadrons in the same event:\newline
        $p_{T,i}\geq0.2~\mathrm{GeV}/c$, $|\eta_i|\leq5$. \\

    \midrule
    \multicolumn{2}{@{}l}{\textbf{Model input}} \\[2pt]

    Electron
      & $\mathbf{x}_e=(p_T^e,\eta_e,q_e)$ \\[2pt]

    Hadron $i$
      & $\begin{aligned}[t]
          \mathbf{h}_i = (p_{T,i},\Delta\eta_i,\sin\Delta\phi_i,\cos\Delta\phi_i,q_i)
        \end{aligned}$ \\[3pt]

    Event
      & $\bigl(\mathbf{x}_e,\{\mathbf{h}_i\}_{i=1}^{N_h}\bigr)$;\newline
        unordered hadron set, $N_h\geq0$. \\

    \bottomrule
\end{tabularx}
\end{table}

Since prompt $D$ and direct $B$ electrons have different transverse-momentum spectra, a $p_T^e$-binned downsampling procedure is applied so that their yields are balanced within each $p_T^e$ interval. This reduces the the influence of the different electron $p_T$ spectra.

For the collision-energy dependence study, additional Pythia samples are generated at $\sqrt{s}=5.02$, 7, and 13.6 TeV. The $0.2$-TeV RHIC-energy sample is based on heavy-flavor hard processes ($gg,q\bar{q}\rightarrow c\bar{c},\,b\bar{b}$), whereas the LHC-energy samples are generated using the \texttt{SoftQCD:nonDiffractive} setting~\cite{Skands:Monash2013}. A common mid-rapidity acceptance, $|\eta^e|<1$ and $|\eta^h|<1$, is applied to the samples. The same electron transverse-momentum range and hadron transverse-momentum selection criteria are applied at all four energies, with prompt $D$ and direct $B$ samples balanced in each electron-$p_T$ interval.

\subsection{Task Definition}
\label{sec:task}

The task is formulated as a binary classification problem between prompt $D$ electrons and direct $B$ electrons. Prompt $D$ electrons originate from charm-hadron decays without a bottom-hadron ancestor, while direct $B$ electrons originate directly from the bottom-hadron decays. The classifier is trained using these two classes. Non-prompt $D$ electrons from $H_b\rightarrow D\rightarrow e$ are kept as a separate sample for the classifier-response study in Fig.~\ref{fig:score_DB}.

The model produces two output logits corresponding to the prompt $D$ and direct $B$ hypotheses, denoted as $\mathrm{logit}_D$ and $\mathrm{logit}_B$, respectively. We define the discriminant score
\[
s=\mathrm{logit}_B-\mathrm{logit}_D.
\]

Larger values of $s$ are more direct-$B$-like, while smaller values are more prompt-$D$-like. The score is used for performance evaluation and for studying the physics information encoded in the charged-hadron environment.

\subsection{Model Architecture}
\label{sec:model}

\begin{figure*}[t]
    \centering
    \includegraphics[width=0.92\textwidth]{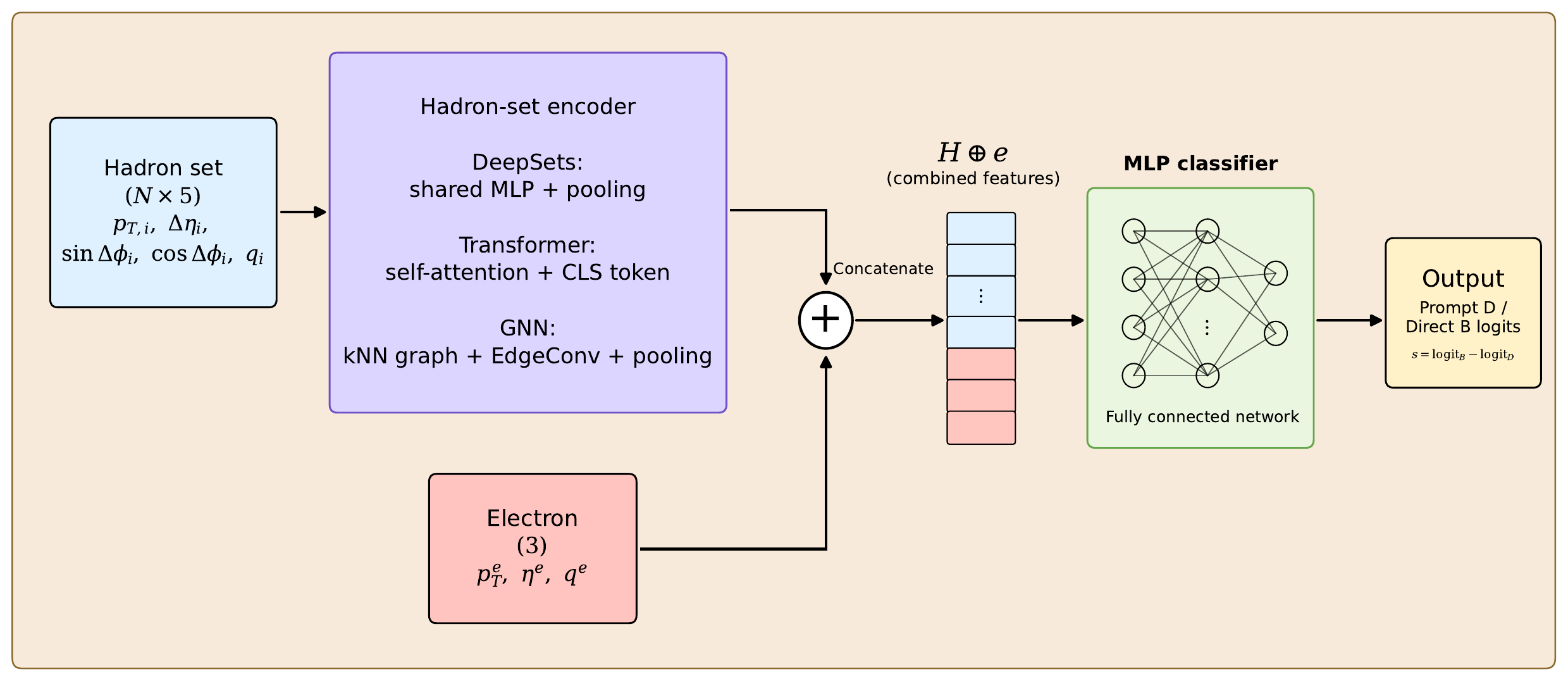}
    \caption{Schematic illustration of the model architecture.}
    \label{fig:model_architecture}
\end{figure*}

The overall model architecture used in this work is illustrated in Fig.~\ref{fig:model_architecture}. The hadronic environment is represented as an unordered point cloud and processed through a set-based learning framework. Different set-modeling architectures are compared within a common classification pipeline.

For each sample, the input hadronic system consists of a variable-length set of charged hadrons characterized by their kinematic and charge information. Each hadron is first processed independently by a shared hadron encoder to obtain a latent particle-level representation.

The resulting set of particle embeddings is then passed to a set-modeling module, which learns correlations and collective structures within the hadronic environment and produces a set-level representation $H$. Three models for set-structured data are considered: a DeepSets model with sum pooling, a four-layer Transformer with a classification (CLS) token, and a graph neural network (GNN) based on a $k$-nearest-neighbor (kNN) graph. For the reported GNN evaluation, the graph is constructed with $k=4$. These architectures provide different mechanisms for modeling interactions among particles and are compared within the same classification pipeline. They are referred to below simply as DeepSets, Transformer, and GNN.

The electron-level features ($p_T^e$, $\eta^e$, and charge) are incorporated together with the hadron-set information within each architecture. The resulting event-level representation is processed by a multilayer perceptron (MLP) classifier to produce the two output logits.

The difference between the logits defines the discriminant score $s$ introduced in Sec.~\ref{sec:task}. This score is used throughout the performance and classifier-response studies.

\section{Results and Discussion}

\subsection{Classification Performance}
\label{sec:classification_performance}

We first evaluate whether the charged-hadron environment contains information that can be used to distinguish prompt $D$ and direct $B$ electrons. The performance is compared using DeepSets, Transformer, and GNN.

\begin{figure}[htb]
    \centering
    \includegraphics[width=\linewidth]{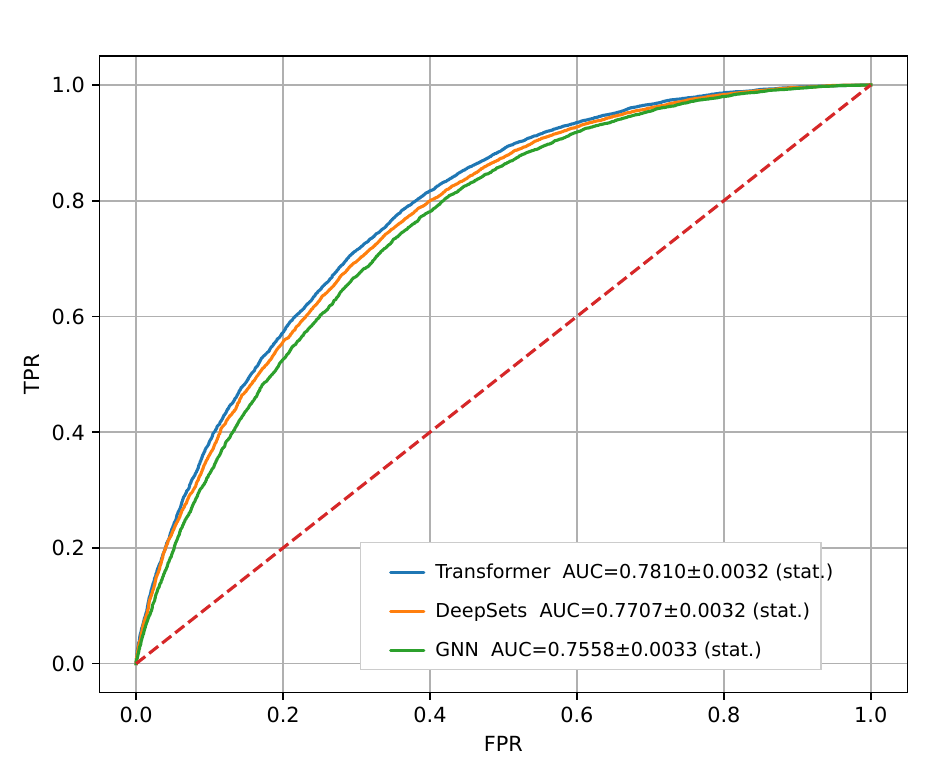}
    \caption{Receiver operating characteristic (ROC) curves for the DeepSets, Transformer, and GNN classifiers. The true-positive rate is the fraction of correctly identified direct $B$ electrons, while the false-positive rate is the fraction of prompt $D$ electrons misidentified as direct $B$ electrons. The legend gives the AUC with its approximate statistical uncertainty.}
    \label{fig:ROC_diff_model}
\end{figure}

The receiver operating characteristic curves are shown in Fig.~\ref{fig:ROC_diff_model}. All three models provide clear discrimination between prompt $D$ and direct $B$ electrons, with Area Under the ROC Curve (AUC) values ranging from about 0.756 to 0.781. Similar classification performance is obtained with the three architectures.

\begin{figure}[htb]
    \centering
    \includegraphics[width=\linewidth]{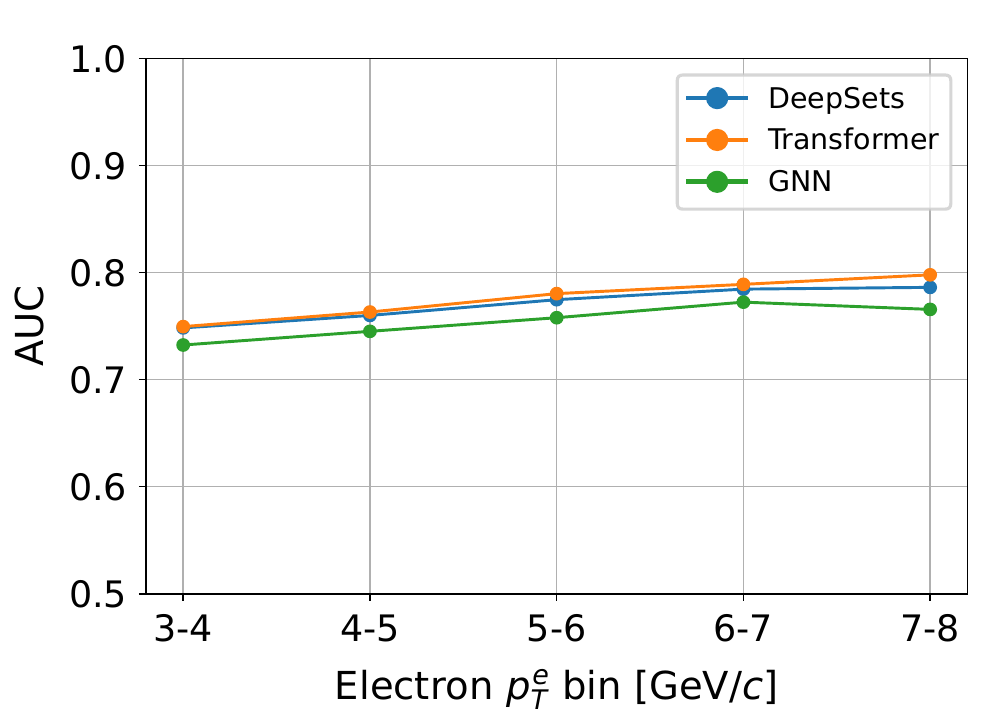}
    \caption{Classification AUC in the plotted electron transverse-momentum range $3<p_T^e<8~\mathrm{GeV}/c$ for the DeepSets, Transformer, and GNN classifiers.}
    \label{fig:AUC_vs_pt}
\end{figure}

The AUC as a function of electron transverse momentum is shown in Fig.~\ref{fig:AUC_vs_pt}. Over the plotted range, the performance improves toward higher $p_T^e$. Similar trends are observed for the three model architectures.

\begin{figure}[htb]
    \centering
    \includegraphics[width=\linewidth]{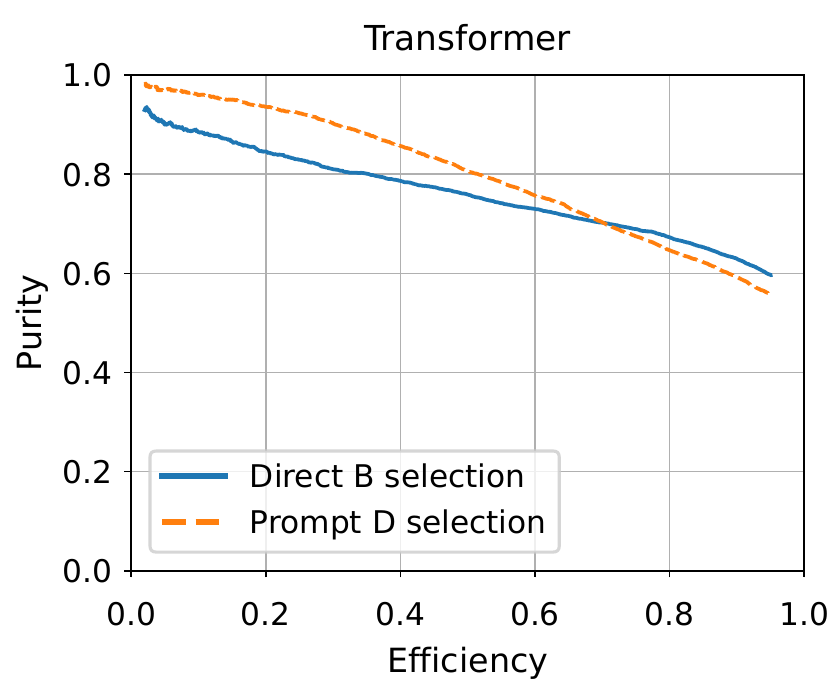}
    \caption{Efficiency--purity performance of the Transformer classifier for direct $B$ and prompt $D$ selections.}
    \label{fig:eff_pur}
\end{figure}

The efficiency--purity relation obtained by scanning the Transformer score threshold is shown in Fig.~\ref{fig:eff_pur}. For both prompt $D$ and direct $B$ selections, the classifier reaches a purity of approximately 80\% at an efficiency of about 40\% on the balanced test sample. The continuous classifier score allows different operating points to be selected according to the requirements of a given analysis.

\subsection{Classifier Response to Physics-Motivated Observables}
\label{sec:physical_interpretability}

\begin{figure}[htb]
    \centering
    \includegraphics[width=\linewidth]{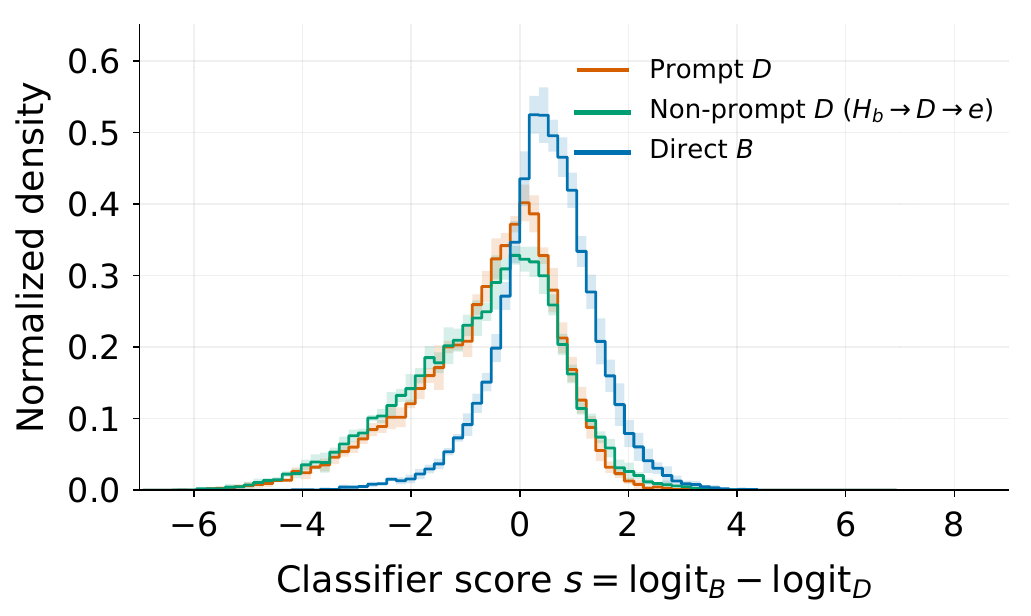}
    \caption{Distribution of the classifier score $s$ for prompt $D$, non-prompt $D$ from $H_b\rightarrow D\rightarrow e$, and direct $B$ electrons. The three samples are matched in electron transverse momentum. The curves and bands show the mean and standard deviation over five Transformer training runs.}
    \label{fig:score_DB}
\end{figure}

Figure~\ref{fig:score_DB} shows the classifier-score distributions for prompt $D$, non-prompt $D$, and direct $B$ electrons. The prompt $D$ and direct $B$ samples form the two training classes, and the trained classifiers are then applied to the non-prompt $D$ sample.

A clear separation is observed between the prompt $D$ and direct $B$ distributions, with substantial overlap between the two classes. The non-prompt $D$ distribution is much closer to the prompt $D$ distribution than to the direct $B$ distribution, although these electrons originate from $H_b\rightarrow D\rightarrow e$ decay chains. This behavior is consistent with sensitivity to hadronic-structure differences associated with the later stages of the decay chain, including the decay of the electron's direct parent hadron.

We next examine the dependence of the Transformer classifier score on several physics-motivated observables.

\begin{figure*}[t]
\centering
\captionsetup[subfigure]{font=footnotesize,labelfont=bf}

\begin{subfigure}{0.48\textwidth}
    \centering
    \includegraphics[width=\linewidth]{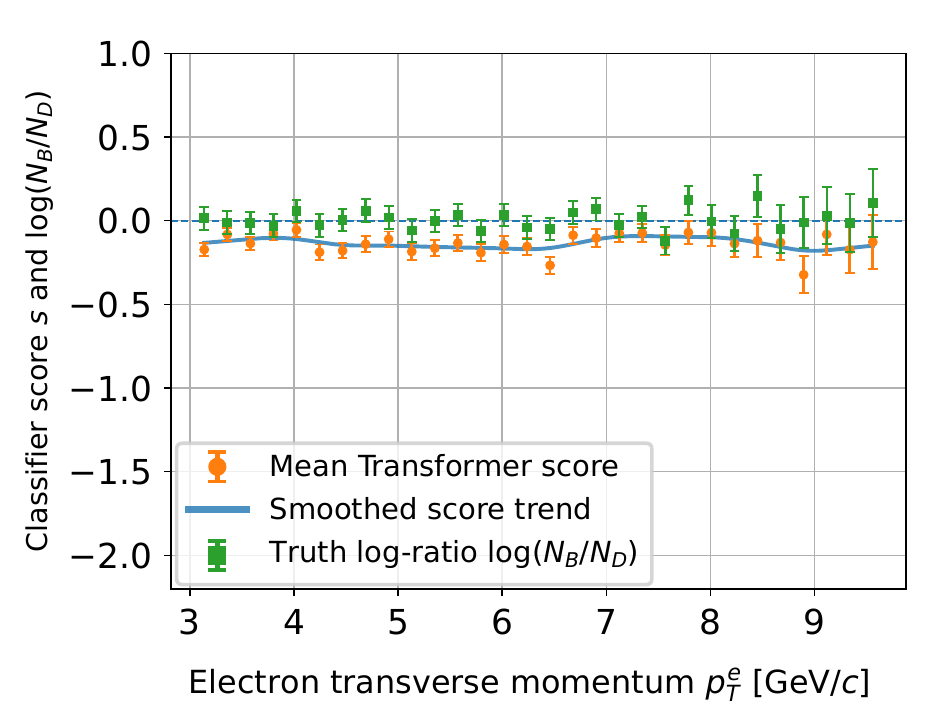}
    \caption{}
    \label{fig:score_dependence_a}
\end{subfigure}
\hfill
\begin{subfigure}{0.48\textwidth}
    \centering
    \includegraphics[width=\linewidth]{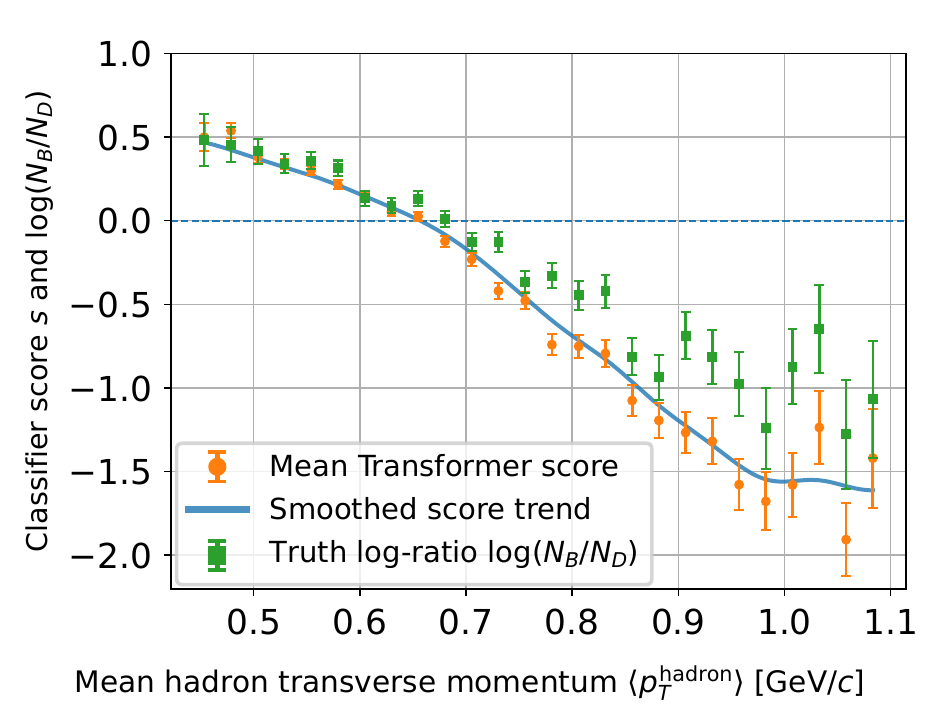}
    \caption{}
    \label{fig:score_dependence_b}
\end{subfigure}

\vspace{0.25cm}

\begin{subfigure}{0.48\textwidth}
    \centering
    \includegraphics[width=\linewidth]{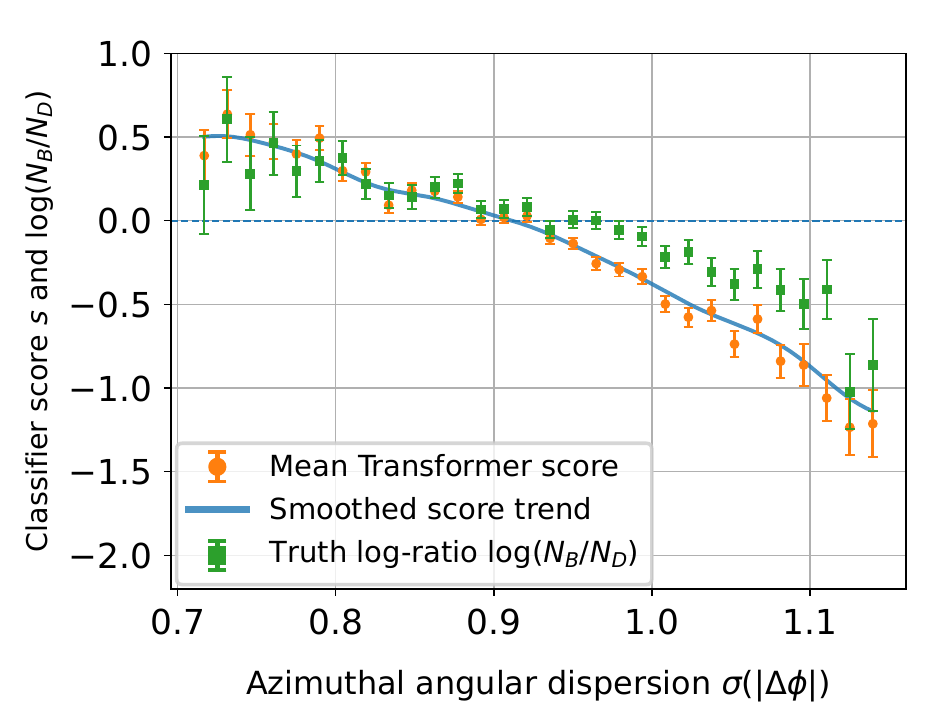}
    \caption{}
    \label{fig:score_dependence_c}
\end{subfigure}
\hfill
\begin{subfigure}{0.48\textwidth}
    \centering
    \includegraphics[width=\linewidth]{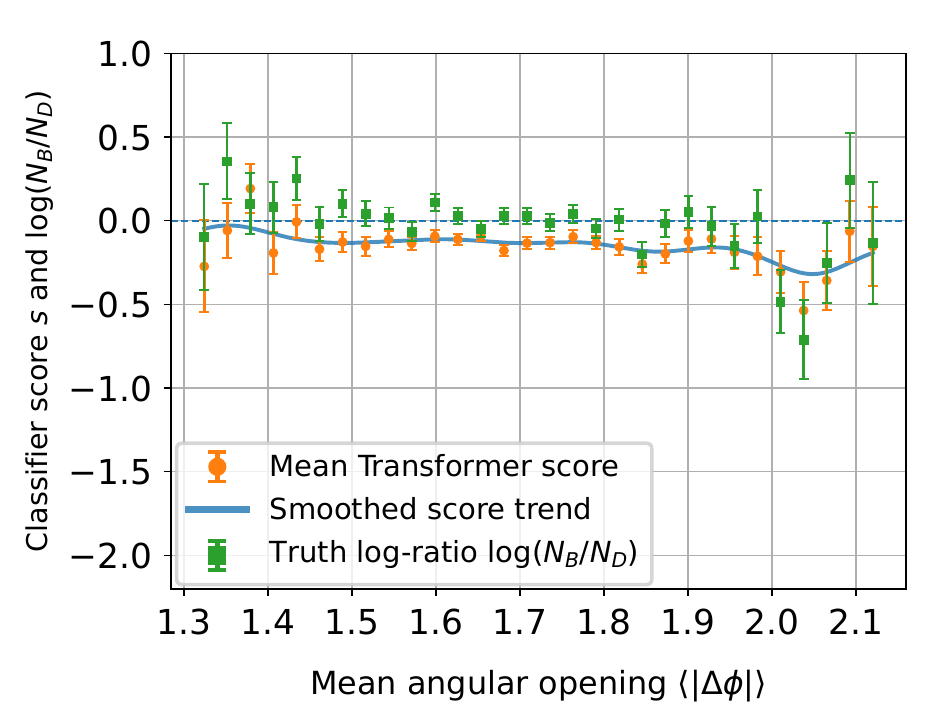}
    \caption{}
    \label{fig:score_dependence_d}
\end{subfigure}

\caption{Dependence of the Transformer classifier score $s$ and the class-yield log-ratio $\log(N_B/N_D)$ on (a) the electron transverse momentum $p_T^e$, (b) the mean hadron transverse momentum $\langle p_T^{\mathrm{hadron}}\rangle$, (c) the azimuthal angular dispersion $\sigma(|\Delta\phi|)$, and (d) the mean angular opening $\langle|\Delta\phi|\rangle$. The points show the mean classifier score in each bin, the line shows its smoothed trend, and the truth log-ratio is shown for comparison.}
\label{fig:score_dependence_scan}
\end{figure*}

As shown in Fig.~\ref{fig:score_dependence_a}, the classifier score exhibits only a weak dependence on the electron transverse momentum. This weak variation is consistent with the $p_T^e$ balancing applied to the two classes.

More pronounced score variations are observed for the properties of the hadronic environment. Figure~\ref{fig:score_dependence_b} shows a clear dependence on the average hadron transverse momentum. The variation with the azimuthal angular dispersion $\sigma(|\Delta\phi|)$ is stronger, as shown in Fig.~\ref{fig:score_dependence_c}: events with broader hadronic angular distributions tend to have smaller scores and are more likely to be classified as prompt $D$ electrons. The dependence on $\langle|\Delta\phi|\rangle$ in Fig.~\ref{fig:score_dependence_d} is weaker.

To compare the learned representation with handcrafted summaries, a boosted decision tree (BDT) is constructed from hadron multiplicity, total, mean, and leading hadron transverse momentum, angular widths in $\eta$ and $\phi$, radial-distance measures, and charge-sum observables.

\begin{figure}[htb]
    \centering
    \includegraphics[width=\linewidth]{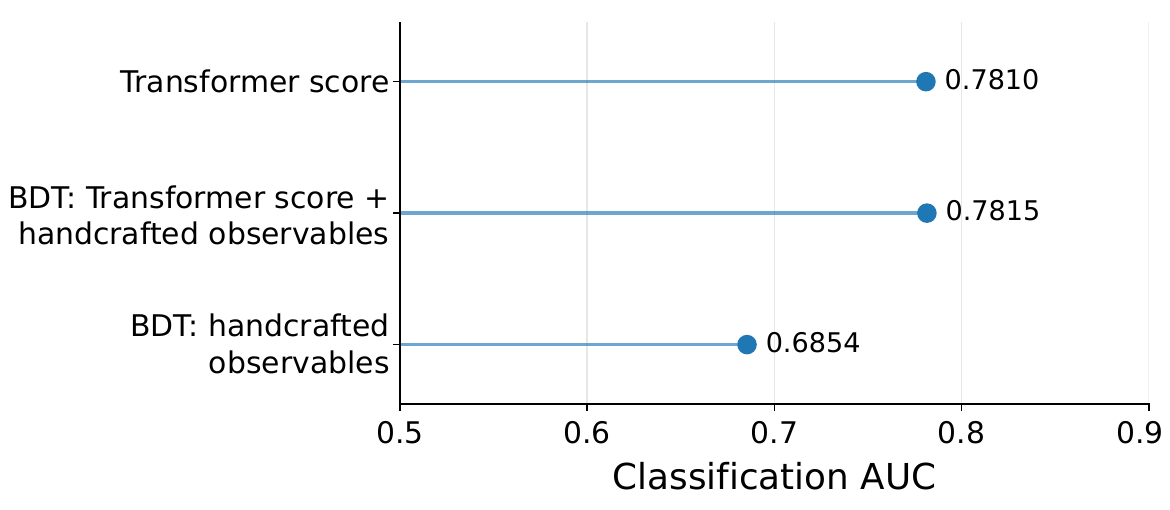}
    \caption{Classification AUC obtained using a BDT based on handcrafted observables, the Transformer score, and a BDT combining the Transformer score with the handcrafted observables.}
    \label{fig:input_compare}
\end{figure}

As shown in Fig.~\ref{fig:input_compare}, the Transformer score gives better classification performance than the BDT based on handcrafted observables. Adding the selected handcrafted observables to the Transformer score produces no visible improvement.

To study how the separation changes across different hadronic environments, events are sorted by the selected observables and divided into ten equal-population percentile intervals. In each interval, the score separation is defined as
\[
\Delta s = \langle s \rangle_{\mathrm{direct\,B}} - \langle s \rangle_{\mathrm{prompt\,D}}.
\]
Larger values of $\Delta s$ correspond to stronger local separation between direct $B$ and prompt $D$ electrons.

\begin{figure}[htb]
    \centering
    \includegraphics[width=\linewidth]{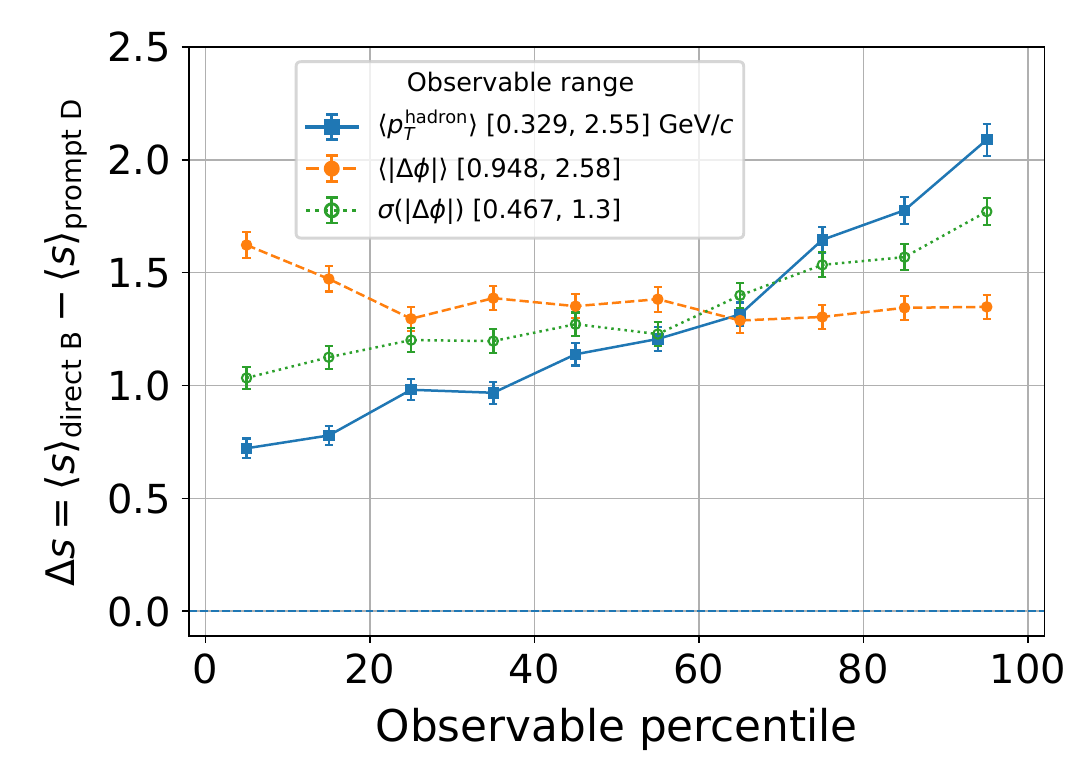}
    \caption{Dependence of the Transformer score separation $\Delta s=\langle s\rangle_{\mathrm{direct\,B}}-\langle s\rangle_{\mathrm{prompt\,D}}$ on equal-population percentile intervals of selected charged-hadron observables. Events are sorted according to each observable and divided into ten equal-population intervals.}
    \label{fig:delta_s_quantile}
\end{figure}

Figure~\ref{fig:delta_s_quantile} shows that $\Delta s$ increases with the average hadron transverse momentum. The separation also varies strongly with the azimuthal angular dispersion $\sigma(|\Delta\phi|)$ and is larger in bins with larger $\sigma(|\Delta\phi|)$. For $\langle|\Delta\phi|\rangle$, the separation is largest at small average angular opening, decreases with increasing angle, and approaches a nearly constant value in the large-angle region.

For the DeepSets classifier, each selected feature is shuffled across events while the remaining inputs are unchanged; the $\sin\Delta\phi$ and $\cos\Delta\phi$ components are shuffled jointly.

\begin{figure}[htb]
    \centering
    \includegraphics[width=\linewidth]{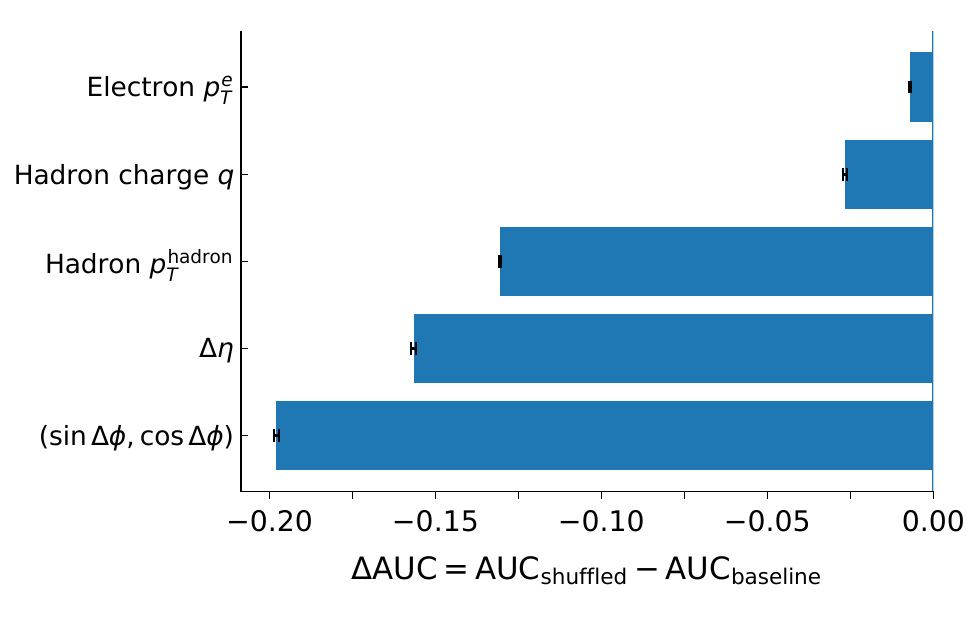}
    \caption{Change in the classification AUC of the DeepSets classifier after event-wise shuffling of selected input features. The $\sin\Delta\phi$ and $\cos\Delta\phi$ components are shuffled jointly.}
    \label{fig:feature_shuffling}
\end{figure}

As shown in Fig.~\ref{fig:feature_shuffling}, shuffling $\Delta\eta$ or the joint $(\sin\Delta\phi,\cos\Delta\phi)$ representation produces the largest decrease in AUC. Shuffling the electron transverse momentum, hadron transverse momentum, or charge has a smaller effect, making the angular variables the most important part of the tested input representation.

The strong response to the angular inputs motivates a direct comparison of the classification performance obtained from different angular regions of the charged-hadron environment.

\subsubsection{Classification with Different Angular Regions}
\label{sec:angular_support}

To study how the discriminating information is distributed around the electron, the Transformer is trained using charged hadrons selected from different angular regions. The distances from the near-side and away-side directions are defined as
\[
\Delta R_{\mathrm{near}}=\sqrt{(\Delta\eta)^2+(\Delta\phi)^2},
\]
\[
\Delta R_{\mathrm{away}}=\sqrt{(\Delta\eta)^2+(\pi-|\Delta\phi|)^2}.
\]
We compare the near-side selection, the combined near- and away-side selection, and the full accepted charged-hadron environment. Cone radii of 0.4 and 1.0 are used for the restricted selections.

For each angular selection, five Transformer classifiers are trained independently. The reported intervals are obtained by event-level bootstrap resampling of the held-out sample with the trained checkpoints fixed, averaging the AUC over the five checkpoints in each replicate.

\begin{figure}[htb]
    \centering
    \includegraphics[width=\linewidth]{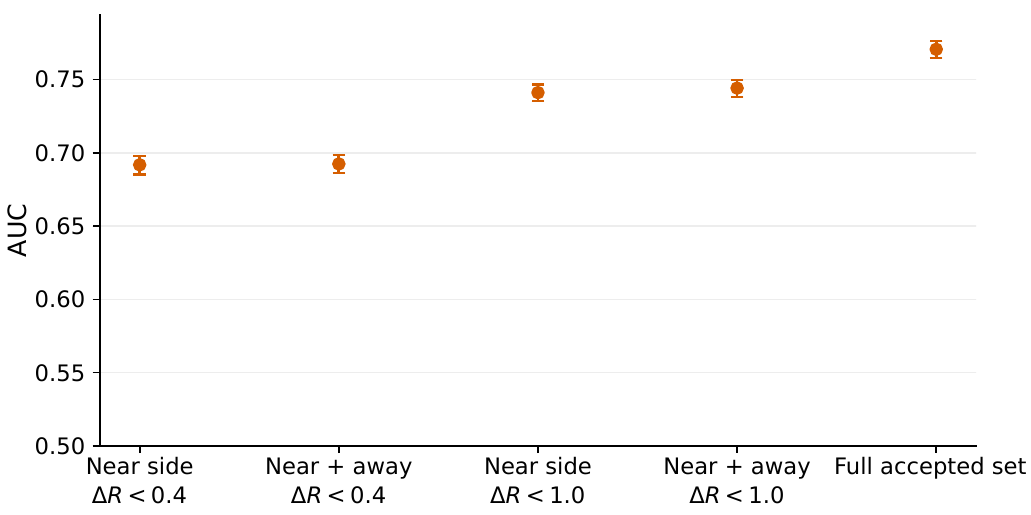}
    \caption{Classification AUC of the Transformer classifiers obtained using charged hadrons from different angular regions. Near-side and combined near- and away-side selections are shown for cone radii of 0.4 and 1.0, together with the full accepted charged-hadron set. The points and error bars show the median and central 95\% interval from event-level bootstrap resampling of the held-out sample with the trained checkpoints fixed.}
    \label{fig:angular_scan}
\end{figure}

As shown in Fig.~\ref{fig:angular_scan}, the restricted angular selections already provide substantial discrimination, and the performance improves as the angular coverage is enlarged. At each tested radius, adding the away-side cone produces only a small change relative to the corresponding near-side selection. The highest performance is obtained using the full charged-hadron environment, indicating that hadrons outside the selected near- and away-side cones also contribute to the separation.

\subsection{Comparison at Different Collision Energies}
\label{sec:collision_energy}

To examine the performance over the collision-energy range relevant to RHIC and the LHC, the same Transformer-based classification method is applied to proton--proton collisions at $\sqrt{s}=0.2$, 5.02, 7, and 13.6 TeV. The comparison uses the common acceptance $|\eta^e|<1$ and $|\eta^h|<1$, with electrons in the range $3<p_T^e<10~\mathrm{GeV}/c$.

\begin{figure}[htb]
    \centering
    \includegraphics[width=\linewidth]{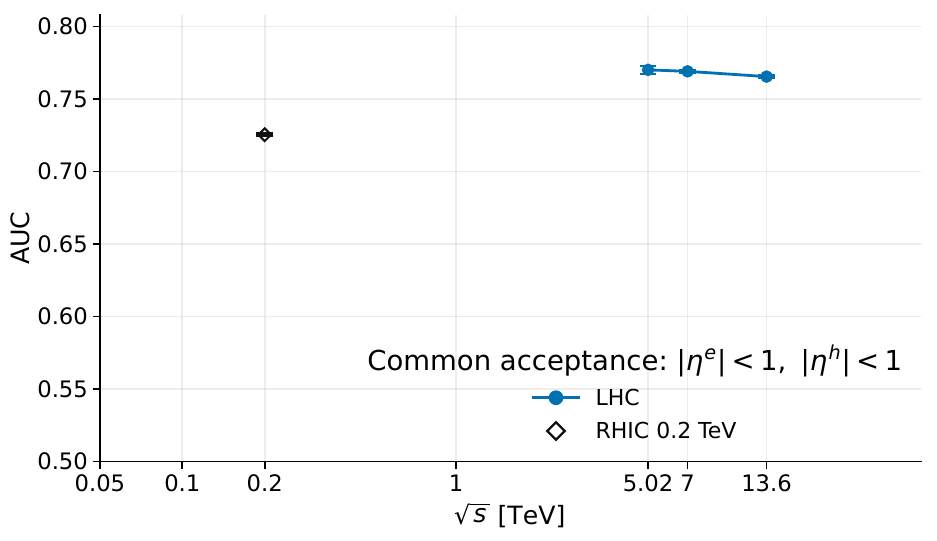}
    \caption{Integrated prompt-$D$/direct-$B$ classification AUC at $\sqrt{s}=0.2$, 5.02, 7, and 13.6 TeV for $3<p_T^e<10~\mathrm{GeV}/c$ under the common acceptance $|\eta^e|<1$ and $|\eta^h|<1$. The markers and error bars show the mean and standard deviation over five Transformer training runs.}
    \label{fig:energy_scan}
\end{figure}

Figure~\ref{fig:energy_scan} shows similar classification performance at the three LHC energies, while the result at $\sqrt{s}=0.2$ TeV is lower. Under the common mid-rapidity acceptance, the studied LHC samples yield higher AUC values than the $0.2$-TeV sample.

\FloatBarrier

\section{Summary}

Separating electrons from prompt $D$ and direct $B$ decays is essential for heavy-flavor measurements, but remains challenging because the two processes exhibit similar decay topologies. In this work, we develop a particle-level approach to separate prompt $D$ and direct $B$ electrons without explicit vertex observables, by representing the charged-hadron environment as an unordered point cloud and combining it with the electron kinematic information. This formulation allows the multi-particle environment to be learned directly without reducing it to a set of composite observables. DeepSets, Transformer, and GNN achieve comparable classification performance. The Transformer performs better than a BDT method based on physics-motivated high-level observables, and no visible improvement is obtained when those observables are added to the score. At an efficiency of about 40\%, the Transformer reaches a purity level close to 80\% on the balanced test sample, demonstrating that substantial separation is achievable using the charged-hadron environment alone.

The classifier response and feature perturbation studies show that the learned separation is particularly sensitive to the momentum and angular structure of the charged-hadron environment. The non-prompt $D$ score distribution is much closer to the prompt $D$ distribution than to the direct $B$ distribution, indicating that the learned separation is driven mainly by the hadronic-structure associated with the later stages of the decay chain rather than by initial production effects. The angular-region study further shows that the charged-hadron environment within the full acceptance gives the highest performance and useful discriminating information is not confined to the selected near- and away-side cones. These findings suggest that the underlying event carries important information.

The method is evaluated for mid-rapidity proton--proton events at both RHIC and LHC energies. The three LHC energies give comparable classification performance, while the result at $\sqrt{s}=0.2$ TeV is lower for the prompt-$D$ and direct-$B$ electron samples in our study. Our results suggest that charged-hadron environments, learned directly from particle-level data, provide an important complementary handle for heavy-flavor electron separation in current and future RHIC and LHC analyses.

\bmhead{Acknowledgements}
We are thankful for support from the National Key R\&D Program of China under Grant No.~2024YFA1610702 and from the National Natural Science Foundation of China under Grant No.~12475187. The computations in this research were performed using the CFFF platform of Fudan University.

\section*{Data Availability}

The datasets generated and analysed during the current study are available from the corresponding author upon reasonable request.

\bibliography{ref}

\end{document}